\documentclass[12pt]{article}
\usepackage{graphics}
\usepackage{epsfig}
\usepackage{latexsym}
\newcommand{\beq}{\begin{equation}}
\newcommand{\eeq}{\end{equation}}
\newcommand{\beqa}{\begin{eqnarray}}
\newcommand{\eeqa}{\end{eqnarray}}
\newcommand{\beqar}{\begin{eqnarray*}}
\newcommand{\eeqar}{\end{eqnarray*}}
\newcommand{\al}{\alpha}

\newcommand{\del}{\delta}
\newcommand{\dee}{{\mit \Delta}}
\newcommand{\w}{\omega}

\newcommand{\ssc}{\scriptscriptstyle}
\newcommand{\eg}{{\it e.g.,}\ }
\newcommand{\ie}{{\it i.e.,}\ }
\newcommand{\labell}[1]{\label{#1}} 
\newcommand{\reef}[1]{(\ref{#1})}

\newcommand\cO{{\cal O}}

\def\IR{{\hbox{{\rm I}\kern-.2em\hbox{\rm R}}}}
\begin{document}

\thispagestyle{empty}
\rightline{\small hep-th/9905081 \hfill McGill/99-10}
\vspace*{2cm}

\begin{center}
{\bf \LARGE Exotic Scalar States }\\[.25em]
{\bf \LARGE in the AdS/CFT Correspondence}
\vspace*{1cm}

Neil R. Constable\footnote{E-mail: constabl@hep.physics.mcgill.ca} and Robert
 C. Myers\footnote{E-mail: rcm@hep.physics.mcgill.ca}\\
\vspace*{0.2cm}
{\it Department of Physics, McGill University}\\
{\it Montr\'eal, QC, H3A 2T8, Canada}\\
\vspace{2cm} ABSTRACT
\end{center}
We investigate a family of solutions of Type IIb supergravity which
asymptotically approach $AdS_{5}\times S^5$ but contain a non-constant
dilaton and volume scalar for the five-sphere. These solutions
preserve an $SO(1,3)\times SO(6)$ symmetry. We discuss the solution
in the context of the AdS/CFT correspondence, and we find that as well as
running coupling from the nontrivial dilaton, the corresponding field
theory has no supersymmetry and displays confinement at least for a
certain range of parameters.
\vfill \setcounter{page}{0} \setcounter{footnote}{0}
\newpage

\section{Introduction} \label{intro}

One of the most interesting aspects of the AdS/CFT
correspondence\cite{juan,gkp,ed} --- for a comprehensive review,
see ref.~\cite{revue} --- is its potential as a theoretical
tool to study real world QCD nonperturbatively\cite{edd,poly}.
In the context of string theory where the AdS/CFT duality is
best understood, simple backgrounds are typically both supersymmetric
and conformally invariant.  In order to make progress 
in investigating real world QCD, it is necessary to construct models in which
both of these symmetries are eliminated. The
initial suggestion\cite{edd} in this direction
was to compactify a higher dimensional theory with nonsupersymmetric
boundary conditions. 
In this way, ordinary four-dimensional Yang-Mills
theory can be studied through an M5-brane construction, and similarly
D3-branes can be used to investigate three-dimensional Yang-Mills
\cite{edd}. This proposal was extensively studied
\cite{ooguri,more,li,gross,mini,sfet3} but there are problems in decoupling
the additional fermions and scalars 
in a regime where the supergravity calculations
can be trusted. Another promising suggestion\cite{zero}
was to study D3-branes in nonsupersymmetric string
theories. Various investigations
\cite{zero,zero2,joe2,mini} have shown that these theories often
reproduce log-scaling of the gauge 
coupling in the ultra-violet. Drawbacks in this case are that 
connecting the UV and IR solutions requires considerable fine
tuning, and even when this can be done the IR solutions are usually
dominated by $\alpha^{\prime}$ corrections.

Another complementary
approach has been to consider new nonsupersymmetric backgrounds
within the context of the usual Type IIb superstring theory. One can
preserve the Poincare invariance of the field theory by exciting scalar
fields in the supergravity background. This corresponds to exciting
scalar operators in the dual Conformal Field Theory (CFT).
 Such constructions have been
presented in refs.~\cite{sfet,gub,extra,added} --- similar supersymmetric
constructions
may be found in ref.~\cite{tsyt} --- and it has been found
that the dual field theories do indeed exhibit QCD-like behavior. In
these examples \cite{sfet,gub,extra,added}, evidence was found for both
confinement and running of the gauge coupling.
In this paper, we continue this line of investigation
with the examination of a new two-parameter family of solutions
of the Type IIb supergravity equations which are asymptotically
$AdS_5\times S^5$ but contain two non-constant scalars: the dilaton
and the volume scalar for the five-sphere. In these solutions, supersymmetry
and conformal invariance are both broken by the presence of the
nontrivial scalars.
They also display interesting QCD-like behavior, a running gauge coupling
and confinement in the infra-red. The latter is signalled by
an area law for Wilson loops and a mass gap in the glueball spectrum,
at least for a certain range of the parameters.
One property that distinguishes the present solutions 
from those considered in refs.~\cite{sfet,gub,extra,added}
is that we can demonstrate that they are realized as the
throat-limits of asymptotically flat D3-brane space-times.

A summary of our key results is as follows.
For our new supergravity solutions, we demonstrate that:
{(i)} The dual field theories exhibit both
confinement in the infrared as well as running of the gauge coupling,
for a certain range of the parameters. {(ii)} Unfortunately, 
only a limited sector of the low-energy physics is insensitive to
the region of strong curvatures at the center of the supergravity solutions.
{(iii)} Further, there is no obvious way to decouple the fermions and scalars
in the dual field theory, essentially because the supergravity solutions
contain a single dimensionful parameter.
{(iv)} The long-range supergravity fields have a standard
interpretation in terms of the expectation values of operators in
the dual field theory. In particular, one finds $\langle T_{ab}\rangle=0$.
{(v)} In conjunction with ({iv}), the nontrivial scalar field
profiles are {\it not} triggered by the irrelevant operator ${\cal O}_8$,
as suggested in refs.~\cite{sfet,gub}. All of these results
should be generic for the Poincare invariant class
of solutions considered in this approach to constructing QCD-like field
theories \cite{sfet,gub,extra,added}.

The remainder of the
paper is organized as follows: In section 2, we describe the new
two-parameter family of solutions of the ten-dimensional supergravity
equations. In section 3, we show that this geometry produces
an area law for Wilson Loops. Section 4 addresses the issue of a mass
gap, and in section 5, we discuss our results and consider the dual field
theory interpretation of these solutions.

\section{Background Solutions} \label{one}

In this paper we will investigate a two-parameter family of
ten-dimensional solutions of type IIb supergravity. The Einstein-frame
metric is given by:  
\beqa 
ds^2&=& H^{-1/2}\left(1+
\frac{2\w^4}{r^4}\right)^{\delta/4}(-dt^2+dx^2+dy^2+dz^2)
\nonumber\\
&&\qquad\qquad+
H^{1/2}\left(1+\frac{2\w^4}{r^4}\right)^{\frac{2-\delta}{4}}
\left[\frac{dr^2}{\left(1+\frac{\w^4}{r^4}\right)^{5/2}}
+\frac{r^2\,d{\Omega}^2_5}{\left({1+\frac{\w^4}{r^4}}
\right)^{1/2}}\right]
\labell{ds}\\ 
&&\quad{\rm where} \;\;\;H=\left(1+\frac{2\w^4}{r^4}\right)^{\delta}-1
\nonumber
\eeqa 
These solutions are parameterized by two constants: $\w$ which has
the dimensions of length, and $\delta$ which is dimensionless.
We will assume without loss of generality that $\w$ is positive, and
 requiring the metric to be real requires that $\del$ is
also positive. The dilaton is given by:
\beq
e^{2\phi}=e^{2\phi_{0}}\left(1+\frac{2\w^4}{r^4}\right)^{\dee} 
\labell{dilaton}
\eeq 
where $\phi_{0}$ is an arbitrary constant, and the exponent $\dee$
must satisfy 
\beq
\dee^2+\delta^2=10\ .
\labell{expone}
\eeq
The ``electric'' components of the RR five-form are determined
by the potential
\beq
A^{(4)}_{\rm elec}=-{1\over4\kappa}H^{-1}\ dt\,dx\,dy\,dz
\labell{ramram}
\eeq
where $\kappa^2=8\pi G$ 
with $G$ being the ten-dimensional Newton's constant.
The remaining ``magnetic'' components are determined by
requiring $F^{(5)}=*F^{(5)}$. One can express the result as
\beq
F^{(5)}={2\del\w^4\over\kappa}\left( \varepsilon(S^5) + *\varepsilon(S^5)
\right)
\labell{fivef}
\eeq
where $\varepsilon(S^5)$ is the volume form on a round five-sphere
of unit radius.

In the following, it will also be useful to have
the string-frame metric, which is given by: $G_{MN}=e^{(\phi-\phi_0)/ 2}
g_{MN}$ where $G_{MN}$ and $g_{MN}$ are the
string-frame and Einstein-frame metrics, respectively. Hence
the metric in the string-frame becomes
\beqa
 dS^2 &=& H^{-1/2}\left(1+
\frac{2\w^4}{r^4}\right)^{\frac{\dee+\delta}{4}}(-dt^2+dx^2+dy^2+dz^2)
\nonumber\\
&&\qquad\qquad
+H^{1/2}\left(1+\frac{2\w^4}{r^4}\right)^{\frac{\dee-\delta
+2}{4}}\left(\frac{dr^2}{\left(1+\frac{\w^4}{r^4}\right)^{5/2}}
+\frac{r^2\,d{\Omega}_5^2}
{\left(1+\frac{\w^4}{r^4}\right)^{1/2}}\right)
\labell{stds}
\eeqa

Asymptotically at large $r$, the geometry approaches  $AdS_5\times
S^5$ where the radius of curvature of both the AdS factor and the
sphere is given by
\beq
L^4=2\delta \w^4\ .
\labell{length}
\eeq
The usual $AdS_5\times S^5$ solution can be reproduced as a (singular)
limit: set $\del=1$ and then take $\w\rightarrow0$, while rescaling the
boundary coordinates $(t,x,y,z)$ and the inverse string tension $\alpha'$.

In either the Einstein-frame \reef{ds} or string-frame \reef{stds},
the surface $r=0$ is time-like, \ie the tangent space contains a time-like
direction. We also note that proper distance is finite in going from $r=0$
to finite $r$ along a radial line. Investigating curvature invariants
such as $R$ or $R_{KLMN}R^{KLMN}$, one finds that they diverge as
$r\rightarrow0$, and hence $r=0$ is a naked singularity. Again,
this result applies for both the Einstein- and string-frame metrics.
In the latter case, the divergence of the curvature invariants signals
the breakdown of the $\alpha'$ expansion, and so one can not trust the
precise form of the solution in this vicinity
as a string theory background. One can also see the singularity
at $r=0$ by checking simpler invariants involving the dilaton field.
For example, as $r\rightarrow0$,
\beq
\nabla_M\phi\ \nabla^M\!\phi\ \longrightarrow\ 
{\dee^2\over2\w^2}\left({2\w^4\over r^4}\right)^{10-\del-\dee\over4}
\labell{dila2}
\eeq
using the string-frame metric \reef{stds}. Setting $\dee=0$ in
the exponent above yields
the result for the Einstein-frame \reef{ds}. This expression
diverges for any of the allowed values of $\del$, independent of
the sign of $\dee$.  This divergence also indicates the breakdown of the
$\alpha'$ expansion near $r=0$. 
One finds that the Ricci scalar displays precisely
the same divergent behavior as $r\rightarrow0$.
Of course, the result in eq.~\reef{dila2} vanishes for $\dee=0$
(\ie $\del=\sqrt{10}$) since the dilaton \reef{dilaton} is constant.

Turning to eq.~\reef{dilaton}, one sees that for positive $\dee$
the string coupling $e^\phi$ diverges as $r\rightarrow 0$.
Hence in this case, the string loop expansion is uncontrolled near
$r=0$, providing an additional reason that the detailed form of solution
can not be trusted in this vicinity.
On the other hand, for negative $\dee$, the string coupling vanishes as
$r\rightarrow0$.

The new solutions break all of the supersymmetries as
can be seen by the transformation of the supergravity spinors\cite{john}
\beq
\delta\lambda={1\over2}\gamma^M\partial_M\phi\, \epsilon^*
={1\over2}\gamma^r\partial_r\phi \,\epsilon^*\ .
\labell{super}
\eeq
Hence there can be no Killing spinor solutions for
$\delta\lambda=0$. Of course, this variation vanishes in
the case of a constant dilaton (\ie $\dee=0$), and so one would
have to consider the gravitino variation.
We will not pursue this case further here.

Given the pathological behavior at $r=0$,
a complete understanding of these solutions
and the dual field theory requires a full knowledge of string
theory. In many cases, however, we will be able to extract the
interesting physics while only considering regions where supergravity is
reliable. We can trust the supergravity approximation as long as curvature 
scales are large compared to the string scale, and the string coupling is
small. As a prerequisite then, we will require in the
asymptotic region
\beq
\frac{L^4}{\alpha^{\prime 2}}={2\del\w^4\over\alpha^{\prime 2}}
\gg  1 \;\;\;{\rm and}\;\;\;
e^{\phi_{0}}\ll 1\ .
\labell{limits}
\eeq
Recall that the former quantity is related to the
't Hooft coupling
\beq
\lambda= g_{\ssc YM} \sqrt{N}=\frac{L^2}{\alpha^{\prime}}\ .
\labell{coupling}
\eeq
in the ultraviolet regime for the $U(N)$ gauge theory.

Since the present family of solutions differ from $AdS_5\times
 S^5$ by virtue of nontrivial excitations of the dilaton, 
the volume scalar for the five-sphere
and components of the five-dimensional metric, one expects 
that in the dual ${\cal N}=4$ super-Yang-Mills theory, expectation 
values for various operators, such as Tr($F^2$), have been turned on. From 
the field theory point of view, these expectation values break the conformal 
invariance and the supersymmetry. This in turn produces a running of the gauge
coupling and confinement in the infra-red, as we will demonstrate below.

\section{Quark-Antiquark Potential}

The first test of confinement that we carry out is the calculation
 of the quark-antiquark potential\cite{jaun2,rey2,itz,gross}. The standard 
procedure is to minimize the Nambu-Goto action for a fundamental test 
string in the background given above. The action is,
\beq
S = \frac{1}{2\pi\alpha^{\prime}}\int d^2\sigma\sqrt{-\det G_{MN}
\,\partial_ax^M\partial_bx^N}
\labell{action}
\eeq
where $G_{MN}$ denotes the string frame metric~\reef{stds}. 

We are interested in finding a static solution to the equations of
 motion so it is convenient to set $\sigma^0=t$ and $\sigma^1=x$, and
 assume $y=z=0$. With these choices, the action becomes
\beq
S = \frac{1}{2\pi\alpha^{\prime}}\int dtdx\;n(r)\left[{1+
m(r)\left(\frac{dr}{dx}\right)^2}\right]^{1/2}
\labell{index}
\eeq
with
\beqa
n(r) &=& H^{-\frac{1}{2}}\left(1+\frac{2\w^4}{r^4}\right)^{\frac{\delta
+\dee}{4}} 
\labell{nnrr}\\
m(r) &=& H\frac{\left(1+\frac{2\w^4}{r^4}\right)^{1-\delta\over 2}}
{\left(1+\frac{\w^4}{r^4}\right)^{5\over2}}
\nonumber
\eeqa
If $n(r)$ possesses a global minimum at some position $r_{min}$ then
a long string situated at this minimum is a solution of the equations
of motion. It is straightforward to show that
$n(r)$ does have a global minimum located at
\beq
r_{min}^4= \frac{2\w^4}{(\frac{|\dee| + \delta}{|\dee|
-\delta})^{1\over{\delta}}-1}
\labell{rmin}
\eeq
for $\dee>0$ and $\del \le\sqrt{5}$. (Above, we have added the absolute
values in this formula for later considerations.)
The idea here is that when the end-points of the string
are widely separated at the boundary of the AdS space, the string will
fall into the interior of the space and settle at $r_{min}$ \cite{gub}.
So evaluating the action for large separations, one would find a linear quark
anti-quark potential,
\beqa
V(\Delta x)&\simeq&\frac{n(r_{min})}{2\pi\alpha'}\,\Delta x
= \frac{\Sigma^{\frac{\del-|\dee|}{4\delta}}}
{2\pi\alpha'\sqrt{1-\Sigma}}\,\Delta x
\labell{qqpot}
\\
&{\rm with}& \;\;\;\Sigma=\frac{|\dee|-\delta}{|\dee|+\delta}
\nonumber
\eeqa
where the coordinate distance $\Delta x$ between the string endpoints
corresponds to the separation between the quark and anti-quark in the
background geometry of the dual field theory. Further using
the above definition of the gauge coupling \reef{coupling},
the QCD string tension can be written as
\beq
T_{\ssc QCD}\simeq {n(r_{min})\lambda\over2\pi L^4}\ . 
\labell{tension}
\eeq
The linear quark-antiquark potential is then evidence
for the confinement of electric charge in the dual gauge theory.

\begin{figure}[ht!]
\center{\includegraphics{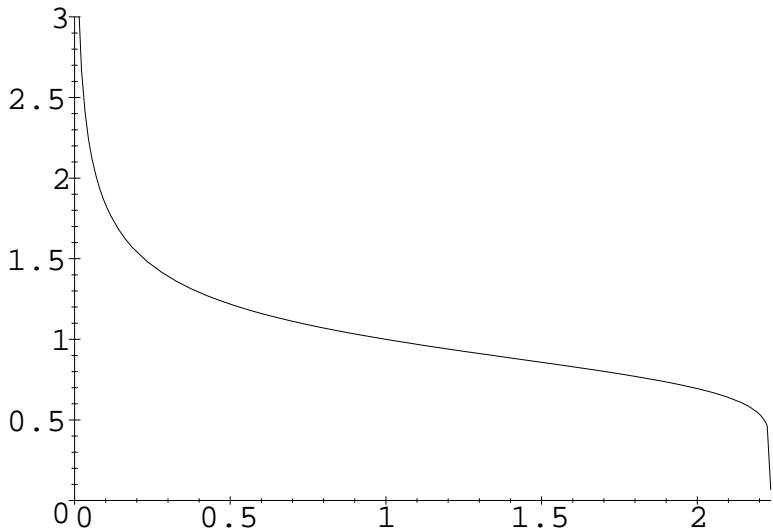}}
\caption{Plot of ${r_{min}}/{L}$ (vertical axis) versus $\delta$. 
Here $L$ is the characteristic curvature scale 
of the asymptotic AdS region, given by eq.~\reef{length}. Note that
as $\delta$ approaches zero, $r_{min}$ is getting large and as 
$\del\rightarrow \sqrt{5}$, $r_{min}\rightarrow 0$.}
\label{fig:rmin}
\end{figure}

\begin{figure}[ht!]
\center{\includegraphics{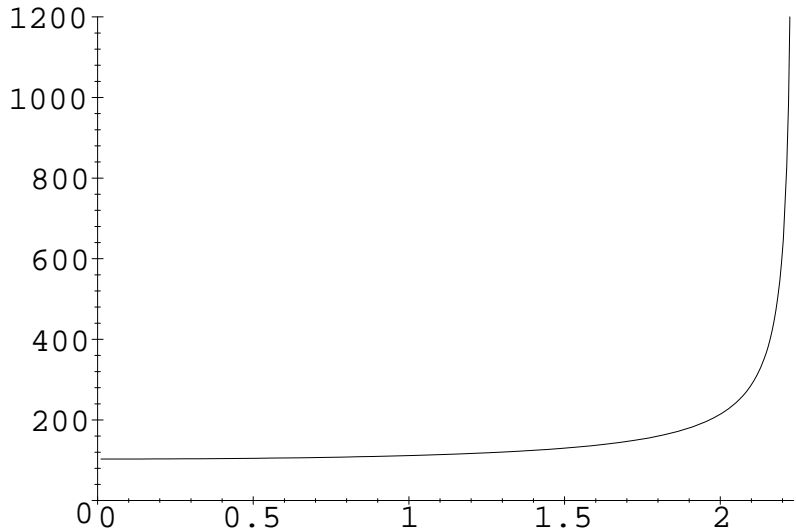}}
\caption{Plot of $L^2(\nabla\phi (r_{min}))^2$ (vertical axis) versus $\delta$.
Here $\phi (r)$ is the background dilaton \reef{dilaton}, and $L$ is the
asymptotic AdS scale \reef{length}.
The figure suggests that the $\alpha^{\prime}$ expansion 
is breaking down at $r_{min}$ for $\delta>2$.
}
\label{fig:Rplot}
\end{figure}

\begin{figure}[ht!]
\center{\includegraphics{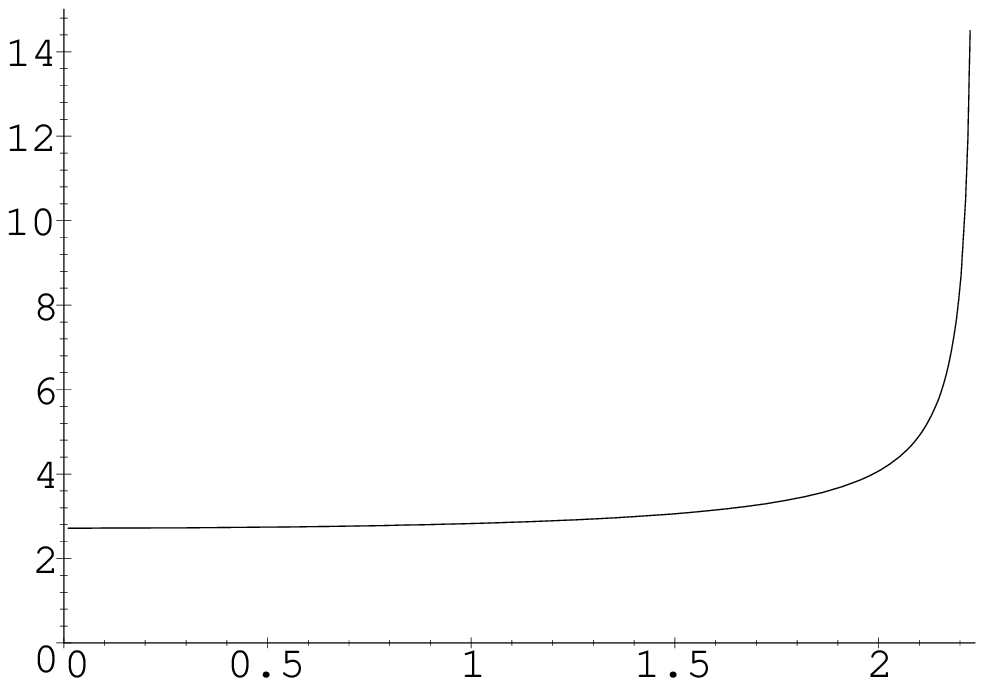}}
\caption{Plot of the $e^{\phi(r_{min}) - \phi_{0}}$ (vertical axis) versus
$\delta$.  The vertical axis gives ratio of the string coupling
evaluated at $r_{min}$ to that in the asymptotic AdS region. 
The plot is made for positive $\dee=\sqrt{10-\del^2}$ in
eq.~\reef{dilaton}. In this case, the figure
suggests that the string loop expansion will become uncontrolled for
$\delta > 2$.}
\label{fig:coupling}
\end{figure}

We should note that the validity of our calculation depends on 
$r_{min}$ being far from the origin since, as discussed above, the
supergravity approximation is not valid in this vicinity. From
eq.~\reef{rmin}, we find that $r_{min}$ is of the order of $\omega$
for small values of $\delta$, and that $r_{min}\rightarrow 0$ as
$\delta\rightarrow\sqrt{5}$. For $\delta>\sqrt{5}$ or $\dee<0$,
$n(r)$ has no minima at finite $r$. Rather it monotonically
decreases to zero at $r=0$ for this parameter regime.
Plotting the ratio of $r_{min}$ and
the asymptotic AdS radius $L$ in fig.~\ref{fig:rmin},
we see that for $\delta<2$ (and $\dee>0$), $r_{min}$ is roughly of the order
of $L$, which is assumed to be large. Hence we expect that
our results are reliable for this parameter range. 
For $\delta>2$ or $\dee<0$, we can not draw any quantitative
conclusions about the
quark-antiquark potential without going beyond the supergravity
approximation. One might observe the trend, however, that the QCD string
tension \reef{tension} is decreasing as $\delta$ approaches $\sqrt{5}$
(and hence, at the same time, approaches $\dee$). Note that curiously
the limiting value is $T_{\ssc QCD}|_{\delta=\sqrt{5}}=1/(2\pi\alpha')$, which is
precisely the fundamental string tension in flat empty space.

We checked the previous expectations
with figs.~\ref{fig:Rplot} and \ref{fig:coupling}.
Fig.~\ref{fig:Rplot} shows that $(\nabla\phi)^2$ is well behaved at $r_{min}$
for $\delta<2$, but that for larger values, the $\alpha^{\prime}$ expansion is
breaking down in the vicinity of $r_{min}$. Hence our results
calculated in the supergravity approximation are not reliable
for these large values of $\delta$.
Similarly, fig.~\ref{fig:coupling} shows that over
the range $0<\del <2$ (with $\dee$ positive) the string coupling at $r_{min}$
is roughly the order of the asymptotic value, which is taken to be small.
However, for larger values of $\delta$, the position of the test string
strays into regions where the string loop expansion is uncontrolled, and
hence the details of the supergravity solution are unreliable.

We might attempt to determine the monopole-antimonopole
potential\cite{mono}
by repeating the above calculation with the action of
a D-string which is given by
\beq
S = \frac{1}{2\pi\alpha^{\prime}}\int d^2\sigma\, e^{-\phi}\sqrt{-\det G_{MN}
\,\partial_ax^M\partial_bx^N}
\eeq
Assuming the static gauge as above, the action has precisely the same
form as in eq.~\reef{index} except that the function $n(r)$ is replaced by
\beq
N(r)= H^{-\frac{1}{2}}\left(1+\frac{2\w^4}{r^4}\right)^{\frac{\delta
-\dee}{4}}\ .
\eeq
Now for $\dee>0$ as in the case of interest above,
$N(r)$ has no extrema for any real values of $r$.
Rather the function $N(r)$ is a monotonically decreasing function of
$r$ as the string approaches the singularity. If one were
to use this result, it would appear that the most energetically favorable
situation is
for the string to fall all the way to the singularity where $N(r)$
vanishes and then to run along the singularity for a distance $\Delta x$
before returning to the asymptotic region of the space-time. This 
is possible because at $r=0$ it costs no energy to separate the 
monopoles and the interpretation is that the magnetic charge is
screened~\cite{joe2}. However as discussed, we can not trust
this conclusion since the supergravity approximation
breaks down in the vicinity of $r=0$.

However, one might also consider the monopole-antimonopole potential
~\cite{joe2,li,gross} for $\dee<0$. In this case, 
$N(r)$ does have a global minimum whose
position is given by eq.~\reef{rmin} for $\del\le\sqrt{5}$,
and so a linear potential
of the form \reef{qqpot} arises for the monopoles in this parameter
range. Further note that in this case, $n(r)$ decreases monotonically to
zero at $r=0$, which is suggestive of screening of electric
charge. Hence the quarks and monopoles have interchanged roles
for $\dee<0$. Mathematically one can see that this result
arises because changing the sign of $\dee$ 
(\ie $\dee\rightarrow-\dee$) precisely
interchanges the form of $n(r)\leftrightarrow N(r)$ and so trades the
fundamental test string action for that of a D-string. This is to be
expected since as can be seen from eq.~\reef{dilaton},
$\dee\rightarrow-\dee$ implements the $SL(2,Z)$ transformation
$e^\phi\rightarrow e^{-\phi}$. Hence one might have expected that the
physics of the quarks and monopoles should be interchanged in the
two complementary parameter ranges of positive and negative $\dee$.

\section{Mass Gap and Glueballs}

As argued by Witten\cite{edd}, a second useful criterion in establishing
confinement in the dual gauge theory is the
presence of a mass gap. To demonstrate the existence of a mass gap
in the present case we begin by calculating the mass spectrum for scalar
glueballs. The prescription for performing such a calculation was originally
given in ref.~\cite{ooguri}. The strategy is to examine the linearized equation
of motion  for a scalar field fluctuation $\eta$ propagating in the
supergravity background, and determine the spectrum of $m^2=-k^2$ where $k$ is
the four-momentum of the scalar field in the boundary directions spanned by
$t$, $x$, $y$ and $z$. Making the ansatz $\eta=h(r)e^{ikx}$,
the equation of motion takes the form of a linear second-order ODE for the
radial profile $h(r)$.
In this section, we will consider two scalars, the axion (\ie the RR scalar)
and the dilaton. The linearized equation of motion for the axion
may be written as
\beq   
e^{-(\phi-\phi_0)/2}\left(\nabla^2_{Einstein} C +2(\partial_{M}\phi)(\partial^{M}C)
\right)=\nabla^2_{string}C=0
\labell{axion}
\eeq
while that for the dilaton is
\beq
\nabla^2_{Einstein}{\varphi} = 0\ .
\labell{vardil}
\eeq
Actually the latter deserves some explanation. Because of the nontrivial 
dilaton profile in the background solution, the linearization of the dilaton
equation of motion ({\it i.e.,} $\nabla^2_{Einstein}\phi=0$) includes
terms involving both the dilaton fluctuation $\varphi$ and the metric 
fluctuation $h_{MN}$. In the Einstein frame, the complete equation is
\beq
\nabla^2\varphi+\nabla^N\left(\left(h_{MN}-\frac{1}{2} g_{MN}h^P_P\right)
\nabla_M\phi\right)=0
\labell{lindil}
\eeq
However the second term in eq.~\reef{lindil} can be eliminated with the
gauge choice
\beq
h_{Mr}-\frac{1}{2}g_{Mr}h^P_P=0
\labell{gauge}
\eeq
in which case the equation reduces to eq.~\reef{vardil}. 
Note that this does not mean that there is no perturbation of the metric
which accompanies the dilaton fluctuation. Certainly there is no gauge 
freedom  which can eliminate $\varphi$ terms from the linearized 
Einstein's equations. However the preceding argument that does show that
for the judicious choice of gauge \reef{gauge} above, analyzing the simple
scalar equation ~\reef{vardil} is sufficient to determine the spectrum of
glueballs and there is no need to solve the linearized Einstein equations.
 
Thus from eq.~\reef{axion},the axion propagates as a free field 
on the string-frame geometry,
while from eq.~\reef{vardil}, the dilaton fluctuation does so on 
the Einstein-frame geometry.
We can use this to our advantage in calculating
the spectra of the corresponding glueballs simultaneously. Notice that
setting the parameter $\dee =0$ reduces the string-frame metric \reef{stds}
to the Einstein-frame metric \reef{ds}. 
Thus we can analyze the axion equation \reef{axion}
and by setting $\dee = 0$ obtain the results for the dilaton for
free!\footnote{It should be emphasized that this is a purely formal
statement. In our solution, choosing $\dee =0$
uniquely fixes the parameter $\delta$ by eq.~\reef{expone},
while in the dilaton equation \reef{vardil}, $\delta$ is to be understood as a 
free parameter.}
Making the ansatz $C=h(r)\,e^{ikx}$, the axion equation \reef{axion} becomes
\beqa
\Box_{String}C &=& \frac{\partial^2h(r)}{\partial r^2}+
f(r)\frac{\partial h(r)}{\partial r}+m^2g(r)h(r)=0
\nonumber
\\
f(r)&=&\frac{5r^4+2\w^4\left(1-4\dee\right)}{r\left(r^4+2\w^4\right)}
\nonumber
\\
g(r)&=& \frac{r^{10}\left(\left(1+\frac{2\w^4}{r^4}\right)^{\delta}-1\right)
\left(1+\frac{2\w^4}{r^4}\right)^{(1-\delta)/2}}
{\left(r^4+\w^4\right)^{5/2}}
\labell{eqnmot}
\eeqa
One can check that the dilaton equation \reef{vardil} agrees 
with \reef{eqnmot} upon setting $\dee = 0$.

Applying the methods of refs.~\cite{mini,sfet3}, we change variables to
$r=\w e^y$ and redefine the field $h(y)=\alpha (y)\psi (y)$ so that
the equation is Schr$\ddot{o}$dinger-like:
\beq
0=-\,\psi(y)^{\prime\prime}  + V(y)\,\psi(y)
\labell{schroe}
\eeq
where the effective potential is given by
\beqa
V(y) &=& -\frac{\alpha(y)^{\prime\prime}}{\alpha(y)}-
f(y)\frac{\alpha(y)^{\prime}}{\alpha(y)}-m^2g(y)
\labell{pot}\\
{\rm with}\ \ \alpha (y)
&=& \left(e^{4y}+2\right)^{-\frac{\dee}{4}-\frac{1}{2}}e^{\dee y}
\nonumber
\eeqa
The primes denote differentiation with respect to $y$.

The potential $V(y)$ has the following asymptotic behavior:
\beqa
V(y\gg 0) &\sim & 4 - a\,\w^2m^2e^{-2y}
\nonumber
\\
V(y\ll 0) &\sim & 3\dee^2 - b\,\w^2m^2e^{2(5-\delta)y}
\labell{asymtot}
\eeqa
where $a$ and $b$ are positive constants.

Notice that naively for $y\ll 0$ the $m^2$ term 
in the potential would diverge to $-\infty$ for $\delta > 5$. 
However, from eq.~\reef{expone}, one has the restriction that
$\delta\le\sqrt{10}$ and hence the potential asymptotically approaches a
positive constant for both $y\rightarrow\pm\infty$. Hence we can
expect that eq.~\reef{schroe} will yield normalizable zero-energy
bound states when the potential is appropriately adjusted by
tuning $m^2$, which appears as a parameter in the former.
We can obtain the glueball spectrum in the WKB approximation
following ref.~\cite{mini}. Our
potential has classical turning points located at
\beqa
y_{+}&=&\frac{1}{2}\log\left(\frac{\w^2m^2\delta}{2}\right)
\nonumber
\\
y_{-}&=&-\frac{1}{2(5-\delta)}\log\left(\frac{2^{\frac{1}{2}(\delta+1)
}\w^2m^2}{\dee^2}\right)
\labell{turn}
\eeqa
Now using the prescription and notation of~\cite{sfet3}, we find that
within the WKB approximation 
\beqa
m^2 &=& \frac{\pi^2}{\w^2\xi (\delta)^2}n\left(n + 1 +
\frac{\dee}{5-\delta}\right) +O(1)
\labell{mass}
\\
\xi(\delta) &=& \int_{-\infty}^{\infty}dy\,\sqrt{g(y)}
\quad{\rm and}\quad n\geq1
\nonumber
\eeqa
where $g(y)$ is given in eq.~\reef{eqnmot}.
Note that $\xi(\delta)$ is a number of order one --- more precisely
it rises monotonically from zero to approximately 2.5 as $\delta$ ranges
from 0 to $\sqrt{10}$. 

This result \reef{mass} is reliable as long as the bound
state wave function $\psi(y)$ has no or little support in the region
near the singularity, \ie $y\rightarrow-\infty$, where the
supergravity approximation breaks down. In
eq.~\reef{turn}, the inner turning point will become large and negative
when $\omega m$ becomes large. From eq.~\reef{mass} and the following
discussion, one sees that the latter will occur for small values of $\delta$,
and so in this case, we cannot reliably conclude that there is a mass gap.
Note that for a generic value of $\delta$, $\omega m$ will also become
large when $n>>\xi(\delta)$. In this case, we can conclude that there
is a mass gap, however, we do not have a reliable calculation of
the mass spectrum for very large values of $n$ --- where ordinarily
the WKB approximation is most reliable. There is also a problem as $\delta
\rightarrow \sqrt{10}$, for which $\dee\rightarrow0$ and hence from
eq.~\reef{turn}, $y_-\rightarrow-\infty$. Thus $\delta \sim \sqrt{10}$
would be another range where we could not reliably conclude that there
is a mass gap.

However, for most values of $\delta$ and either sign of $\dee$, we have
a reliable calculation of the mass spectrum corresponding to the operator 
$Tr(F\wedge F)$, and we may conclude that there is a mass gap in the
spectrum of these $0^{-+}$ glueballs of the dual gauge theory.

Finally we turn to the dilaton. We were to reduce the above analysis
for the axion to that for the dilaton by setting $\dee = 0$. However,
we immediately recognize that there is a problem, in that precisely for this
choice the inner turning point is $y_{-}=-\infty$, \ie $r=0$. Hence the potential
doesn't confine the wave-function away from the singularity, and so
the results in eq.~\reef{mass} are not trustworthy. Hence within
the supergravity approximation, we cannot
determine whether or not there is a mass gap in this case.

As a final comment, we also note that a similar problem arises in
analyzing the spin-two glueballs associated with the linearized
fluctuations of the five-dimensional (Einstein) metric.
In this case it also found
that the effective potential for the metric fluctuations
does not confine the wave-function away from the singularity, and hence
we are again unable to determine if there is a mass gap for these states or not.

\section{Discussion}

In this paper we have described a new two parameter family of solutions to 
Type IIb supergravity which are asymptotic to $AdS_5\times S^5$, but
they contain two non-constant background scalars: the dilaton and the
volume scalar for the five-sphere. These nontrivial scalar fields serve 
to break both the supersymmetry and the conformal invariance of the 
background. However, an $SO(1,3)\times SO(6)$ symmetry is preserved. That
is Lorentz invariance in the boundary directions and rotational invariance
of the five-sphere. Further the supergravity solutions contain a naked
time-like singularity at $r=0$. Despite this singularity, we can still
reliably show within the supergravity approximation,
that at least for a certain range of the
parameters, the field theories dual to these solutions exhibit both
confinement in the infrared as well as running of the gauge coupling.
We now discuss a number of aspects of the physics of these supergravity
solutions and the dual field theories.

\smallskip
\noindent{\bf (i) Gauge theory interpretation:}

It is interesting that the new solutions only contain a single dimensionful
parameter $\w$, while the second free parameter is the dimensionless exponent
$\delta$. Combined these two parameters set the AdS radius as given in eq.
~\reef{length}. Similarly the glueball mass spectrum~\reef{mass} is determined
by $\w$ multiplied by some more complicated function of $\delta$. Below
we will also see that $\w$ and $\delta$ combine to determine the expectation
values of certain operators in the dual gauge theory. Given these expectation 
values, it is natural to expect that masses of the form $f(\delta)/\w$ are
induced for the field theory fermions and scalars, and the QCD-like properties 
would then follow. Note that all of the scales, {\it e.g.,} glueball masses 
and fermion masses, described above are tied to a single dimensionful 
parameter $\w$. This means that there is no natural way to suppress the 
fermions and scalars in the theory. Thus one would expect that there are
fermionic glueballs with masses of the same order as those for the scalar
glueballs in eq.~\reef{mass} {\it i.e.,} one could calculate glueball masses
for the linearized gravitino equations as in section 4. This behaviour can be
contrasted with Witten's proposed model for pure Yang-Mills\cite{edd}.
There the compactification radius introduces a second dimensionful scale 
independent of the AdS radius, and a small compactification radius naturally
induces large fermion masses. In any event, our model does not describe
ordinary nonsupersymmetric Yang-Mills theory in four dimensions, rather it
would be Yang-Mills theory coupled to {\it adjoint} fermions and scalars
with masses. However, it provides an interesting frame work in which to study
confinement in four dimensional gauge theories. 

We showed that for $0< \delta < 2$ and $\dee >0$, the
new solutions give rise to a
linear quark-antiquark potential, which may be taken
as evidence for confinement of electric charge. By flipping
the sign of $\dee$ (which amounts to an $SL(2,Z)$ transformation,
sending $e^\phi\rightarrow e^{-\phi}$),
we obtain a solution for which the monopoles of the dual field theory
are confined.

As further evidence for confinement, we also demonstrated that for
all values of the parameter $\del$, except near $\del\sim0$
or $\del\sim\sqrt{10}$, there is a mass gap for fluctuations of
the RR scalar. In the dual gauge theory, this corresponds to
a mass gap in the spectrum of the $0^{-+}$ glueballs. These calculations
are reliable because in the corresponding wave equation,
the effective potential provides a repulsive barrier which
shields the singularity. The same effect has been observed previously
for certain black hole solutions in string theory\cite{donm}. On the other
hand, no such  barrier appears in the linearized wave equations for 
the dilaton or metric fluctuations. Hence determining the spectrum of
these fields would require understanding how the Type IIb string
theory resolves the singularity (if this actually occurs -- see below).

In ref.~\cite{joe2}, a linear quark-antiquark potential and a discrete 
glueball spectrum was also found for a range of parameters while  
investigating type 0 strings in the limit of small radius. In the 
special case that the bulk tachyon is constant in these solutions
the relevant equations of motion should reduce to those considered here
and in refs.~\cite{sfet}-\cite{added} and so the small radius behavior 
would be the same. 

The running of the gauge coupling is simply a consequence of the
dilaton being non-constant. The Yang-Mills coupling is related to
that in the string theory by $g_{YM}^2=2\pi e^\phi$
\cite{juan}. Further
one has the UV/IR relation \cite{suss,peet} relating supergravity
degrees of freedom at large (small) radius to those in the field
theory at high (low) energy. One can interpret the asymptotic
approach of the dilaton to a constant value, \ie $e^\phi-e^{\phi_0}
\propto 1/r^4$, as the flow of the gauge theory to a UV-stable
fixed point \cite{sfet}. Further this particular form for the asymptotic
behavior of the dilaton, which is dictated by the supergravity equations
of motion, can be translated into a universal result for the gauge
theory beta function \cite{sfet}
\beq
\left.{\partial\beta\over\partial g_{\ssc YM}}\right|_{g_{\ssc YM}=
g^*_{\ssc YM}}=-4
\labell{xyz}
\eeq
where $g^*_{\ssc YM}$ is the value of the gauge coupling at the fixed point.
One can confirm that the higher derivatives of the beta function are
non-universal \cite{sfet}, depending on the parameter $\dee$ for the present
solutions.

Further considerations of universal behavior in asymptotically AdS solutions
with nontrivial scalar fields appear in ref.~\cite{univ}. To make contact
with the discussion there, we must first identify the five-dimensional
Einstein metric $g^{(5)}_{\mu\nu}$ from eq.~\reef{ds} with
\beq
ds^2=e^{-{10\over3}\chi}\,g^{(5)}_{\mu\nu}dx^\mu dx^\nu+
L^2e^{2\chi}\,d\Omega^2_5\ .
\labell{fivedim}
\eeq
where $L$ is the asymptotic radius given in eq.~\reef{length}. 
In accord with the prediction of ref.~\cite{univ}, 
one finds that to leading order near the singularity this metric
takes the form
\beq
g^{(5)}_{\mu\nu}dx^\mu dx^\nu\ \propto\ 
d\rho^2+\left({\rho\over\w}\right)^{1/2}
(-dt^2+dx^2+dy^2+dz^2)\ .
\labell{singfive}
\eeq
where $\rho$ is a new radial coordinate.
The prediction of this universal behavior relied on
being able to ignore any potential terms appearing in the scalar
equations of motion near the singularity, which is valid for
polynomial potentials. In fact, the volume scalar $\chi$ has an
exponential potential, as can be seen from its equation of motion
\cite{gub}
\beq
\nabla^2\chi = {4\over L^2}\left(e^{-16\chi/3}-e^{-40\chi/3}\right)\ .
\labell{eqchi}
\eeq
The authors of ref.~\cite{univ} recognize that an exponential potential could
result in violations of the universal behavior considered there. However,
this does not seem to be the case for the present supergravity solutions.
In fact one can easily show that the potential terms are suppressed as 
$r\rightarrow 0$.
Further, as already discussed, we found a linear quark-antiquark potential
for a ``large'' range of the exponents appearing in our solutions, which
is in accord with the analysis in ref.~\cite{univ}. We did not, however,
find the universal glueball spectrum with a mass gap
arising from their calculations.
The obvious reason for this discrepancy is that their universal
spectrum arises for a minimally coupled massless scalar propagating
on the five-dimensional Einstein metric. In the present case of Type IIb
supergravity, there is not obviously any such scalar. Instead we considered
the RR scalar \reef{axion}, which propagates on the ten-dimensional string
metric, and the dilaton \reef{vardil}, which propagates
on the ten-dimensional Einstein metric. We only found
clear evidence of a mass gap in the former case.

\smallskip
\noindent{\bf (ii) Asymptotically flat supergravity solutions:}

The family of supergravity solutions presented here differ from
those in refs.~\cite{sfet,gub} since the present geometries can be 
realized as the throat limits of D3-brane metrics which are asymptotically
flat. Consider the following asymptotically flat solutions with the
Einstein-frame metric given by
\beqa
ds^2&=&F^{-1/2}\left(-f^{\del-\al_1-\al_2-\al_3}dt^2+f^{\al_1}dx^2+f^{\al_2}dy^2+
f^{\al_3}dz^2\right)
 \nonumber\\
&&\qquad+ F^{1/2}f^{2-\del\over4}\left[{dr^2\over \left(1+{\w^4\over r^4}\right)^{5/2}}+
{r^2\,d\Omega^2_5\over\left(1+{\w^4\over r^4}\right)^{1/2}}\right]
\labell{fullm}\\
&&{\rm where}\qquad F=\left(f^\del-1\right)\beta^2+1
\; {\rm and}\; f=1+{2\w^4\over r^4}\ .
\nonumber
\eeqa
The dilaton is given by
\beqa
e^{2\phi}&=&e^{2\phi_0}\,f^\dee
\labell{fulld}
\eeqa
while the RR five-form has $F^{(5)}_{mag}\propto(\beta^2-1)\,\varepsilon(S^5)$.
Here, $\phi_0$ is an arbitrary constant, while $\beta\ge1$ and the exponents
satisfy
\beq
\dee^2+{5\over2}\del^2-4\del(\al_1+\al_2+\al_3)
+4(\al_1^2+\al_2^2+\al_3^2+\al_1\al_2+\al_1\al_3+\al_2\al_3)=10\ .
\labell{constraint}
\eeq
As before, we will also assume that $\w>0$ and $\del>0$.
One may think of this five-parameter solution as describing a non-extremal
D3-brane dressed with extra scalar ``hair'' associated with exciting the dilaton
field and world-volume components of the metric. Note that as a result of
this hair, the surface $r=0$ is generically singular. An exception to this
rule is the choice $\dee=0$, $\al_1=\al_2=\al_3=1$ and $\del=2$ for
which the solution reduces to the standard near-extremal D3-brane.
The solution \reef{fullm} was constructed
by applying an appropriate set of U-duality transformations \cite{sodd}
to a ``hairy'' vacuum solution of ref.~\cite{cmp}.

The corresponding near-horizon solution is constructed by scaling
\beq
r\rightarrow  r/\beta
\qquad
\omega\rightarrow  \omega/\beta
\labell{scale}
\eeq
with $\beta\rightarrow\infty$. In this decoupling limit, $f\rightarrow f$
and so the only change in the form of the solution is that
$F\rightarrow\beta^2H$ with $H=f^{\delta}-1$ as in eq.~\reef{ds}.
Hence the metric \reef{fullm} becomes $ds^2=\widetilde{ds}^2/\beta$
where
\beqa
\widetilde{ds}^2&=&H^{-1/2}\left(-f^{\del-\al_1-\al_2-\al_3}dt^2+f^{\al_1}dx^2
+f^{\al_2}dy^2+f^{\al_3}dz^2\right)
 \nonumber\\
&&\qquad+ H^{1/2}f^{2-\del\over4}\left[{dr^2\over \left(1+{\w^4\over r^4}
\right)^{5/2}}+
{r^2\,d\Omega^2_5\over\left(1+{\w^4\over r^4}\right)^{1/2}}\right]
\labell{nearm}
\eeqa
while the dilaton \reef{fulld} remains unchanged. 
One may easily verify that in the limit $r\rightarrow
\infty$, eq.~\reef{nearm} approaches the product metric on 
$AdS_5\times S^5$ with the radius of curvature given as in eq.~\reef{length}.
 This solution is an extension of
that studied in the present paper because it is no longer Lorentz invariant
in the world-volume directions. Lorentz invariance is restored by setting
$\al_1=\al_2=\al_3=\del/4$, in which case one
recovers precisely the solution in eq.~\reef{ds}. Similarly
the asymptotically flat extension of the latter solution is produced by
making this choice of exponents in eq.~\reef{fullm}
--- note that eq.~\reef{constraint}
agrees with eq.~\reef{expone} in this case.
The new near-horizon solution may also be regarded as a generalization
of those found in ref.~\cite{russo} by the
addition of the dilaton hair.

\smallskip
\noindent{\bf (iii) Dual interpretation of long-range AdS fields:}

We now return to the dual field theory interpretation of these supergravity
solutions. By examining the precise way in which the
solution approaches $AdS_5\times S^5$ asymptotically, 
one may determine the expectation values of various chiral operators
in the CFT \cite{alba}. In considering solutions of
the linearized equations of motion around AdS space, 
one can classify the solutions as
modes which are nonsingular on the interior of AdS,
and those which are nonsingular at the boundary.
For example \cite{ed}, a massive scalar field satisfying
$\nabla^2\,\del\psi=m^2\,\del\psi$ has asymptotic modes
\beq
\del\psi_\pm\simeq \left({r\over L}\right)^{\lambda_\pm} \del\psi_{0\pm}(x^a)
\qquad{\rm with}\ \lambda_\pm=-2\pm\sqrt{4+m^2L^2}
\labell{mode}
\eeq
for asymptotically horospheric coordinates
on $AdS_5$. The modes $\del\psi_+$ extend smoothly over the interior
of the anti-de Sitter space, but produce singularities at the boundary. 
The functions $\del\psi_{0+}$ may be associated with source currents for
the CFT in calculating correlation functions \cite{ed}.
 The modes $\del\psi_-$ are nonsingular
at the AdS boundary, but extend to solutions which are singular
on the interior of the space. The functions $\del\psi_{0-}$ then yield the
corresponding
expectation values, as may be seen from test probes moving
in the supergravity spacetime \cite{alba,scan}. 
As stressed in ref.~\cite{stressa},
the fact that these modes become singular (\ie reach large values)
in the interior of AdS means that one must go beyond the
linearized equations of motion to consider finite expectation values. 
A good example of this behavior are AdS black hole solutions \cite{black}.
These are solutions of the full nonlinear supergravity equations
of motion, which asymptotically approach
AdS space. One can regard the deviations of the metric from
the AdS solution as solutions of the linearized gravity equations, and
a closer examination shows that these linearized solutions correspond
to modes which become singular in the interior of AdS, which
simply indicates that the full solutions enter a nonlinear regime.

In the solutions of interest here, there are essentially three excitations
in the asymptotically AdS solution: the dilaton, the volume scalar $\chi$
and the metric. The massless dilaton couples to the dimension four operator
$\cO_4$,
which is the supersymmetric extension of $Tr(F^2)$. In eq.~\reef{dilaton}
or \reef{fulld}, one finds that asymptotically the deviation from
its asymptotic value gives $\delta\phi\propto r^{-4}$, in accord with
$\delta\psi_-$ in eq.~\reef{mode} for $m^2=0$. Hence there is a corresponding
expectation value in the dual field theory,
\beq
\langle\, \cO_4\,\rangle\  \propto\ \dee \,\w^4\ .
\labell{trf2}
\eeq

{}From eq.~\reef{eqchi}, one can confirm that $m^2L^2=32$ for the volume scalar
$\chi$. This field couples to the dimension eight operator $\cO_8$,
which is the
supersymmetric extension of $STr[F_{a}{}^bF_b{}^cF_c{}^dF_d{}^a-
(F_a{}^bF_b{}^a)^2]$
where the $STr$ indicates that the $U(N)$ generators associated with each
of the field strengths is symmetrized under the trace \cite{ark}.
One may extract $\chi$ from either the original metric \reef{ds}
or its generalization in eq.~\reef{nearm} using eq.~\reef{fivedim}.
An asymptotic expansion of the result confirms that
$\delta\chi\propto r^{-8}$ in accord with
eq.~\reef{mode}. The corresponding expectation value is
then
\beq
\langle\, \cO_8\,\rangle
\ \propto\ (\del^2-4)\,\w^8
\labell{strange}
\eeq

Finally one may calculate the stress-energy tensor using one of the prescriptions
of ref.~\cite{stressa} or \cite{stress}. The deviation of the asymptotic
metric from AdS space in ``nice'' coordinates \cite{stressa} yields
$\delta g_{ab}\propto r^{-2}$, and the resulting stress tensor for the
general solution \reef{nearm} is
\beq
\langle\, T_{ab}\,\rangle=\pmatrix{4(\al_1+\al_2+\al_3)-3\del&0&0&0\cr
                                   0&4\al_1-\del  &0&0\cr
                                   0&0&4\al_2-\del  &0\cr
                                   0&0&0&4\al_3-\del  \cr}\w^4
\labell{noenergy}
\eeq
Note that this result is traceless in general, \ie $\eta^{ab}\langle T_{ab}
\rangle=0$,
and for the Lorentz invariant solution \reef{ds} with
$\al_1=\al_2=\al_3=\del/4$
the stress-energy vanishes completely!

Note that the later seems to be a generic feature of for asymptotically
AdS solutions which preserve the $SO(1,p)$ Lorentz symmetry. By the Lorentz 
symmetry one must have $\langle T_{ab}\rangle=\lambda\eta_{ab}$ for some
constant $\lambda$. This ansatz yields $\langle T^a_a \rangle =\lambda (p+2)$,
however, by the Weyl invariance of the ``boundary'' field theory, one should
also have $\langle T^a_a \rangle =0$ and hence the stress energy must vanish.
We also have explicitly calculated that $\langle T_{ab}\rangle=0$ for the 
solutions presented in refs.~\cite{sfet,gub,extra,added,tsyt}.

\smallskip
\noindent{\bf (iv) The role of $\cO_8$:}

In refs.~\cite{sfet,gub} it was suggested that the nontrivial dilaton
profile in the solutions constructed there may be triggered by considering
the modified theory in which
 $\cO_8$ is added to the ${\cal N}=4$ super-Yang-Mills Lagrangian. We
do not think that this scenario applies for the dual field
theory of the present solutions or those considered in ref.~\cite{gub}.
Despite the appearance of a nontrivial dilaton and volume scalar
in the present solutions, asymptotically these fields only involve
the modes $\psi_-$ and in accord with the standard interpretation,
these fields are dual to expectation values of the corresponding
operators. Having introduced this scale in the field theory 
state, it is natural that fermion and scalar masses would be induced through
loop effects. Hence the QCD-like behavior uncovered in our analysis would be a
natural consequence.
Turning on a microscopic coupling {\it ie.,a constant source current} for $\cO_8$,
 would correspond to exciting in $\chi$ the divergent mode $\del\chi_+
\propto (r/L)^4$, which is absent in the asymptotic expansion
of our solutions. The same is, of course, true for the supergravity
solutions in ref.~\cite{gub}. 

In general, the modes $\del\psi_-$ decay at the boundary and hence
one can expect that the supergravity solutions will remain asymptotically
AdS when these modes are excited. That is introducing a finite scale through 
an expectation value does not greatly disturb the CFT in the UV regime.
The same would be true of the modes
$\del\psi_+$ when $m^2<0$ which can consistently arise in AdS
theories \cite{freetown}. These supergravity fields are dual to
relevant operators in the CFT \cite{ed}, which again have a negligible effect in
the UV regime. Interesting solutions and their interpretation in terms of
renormalization group flows have recently appeared in the literature
\cite{renorm}. For $m^2>0$, $\del\psi_+$ is asymptotically divergent and
so one can expect that introducing a finite excitation by this mode
would destroy the asymptotic AdS structure of the supergravity solution.
On the field theory side, this corresponds to introducing an irrelevant
operator, which is then not suppressed in the UV regime.

We comment on this because for the volume scalar, we can understand
how this occurs, \ie how $\delta\chi_+$ destroys or modifies the
asymptotic structure of the supergravity solution. Let us begin with
the full metric \reef{fullm} and only perform the scaling limit
\reef{scale} with large but finite $\beta$. That is we do not fully
implement the decoupling limit described above. Examining
the resulting solution in a regime where $\beta>>(r^2/L^2)>>1$,
one finds that the geometry is essentially $AdS_5\times S^5$ up to
small perturbations. That is there is a scaling regime in which 
the dual theory behaves like a conformal field theory.
In particular, examining the volume scalar there
is an additional perturbation
\beq
\delta\chi={1\over4\beta^2}{r^4\over L^4}
\labell{new}
\eeq
which corresponds precisely to the mode $\del\chi_+$ given that
$m^2L^2=32$ for this scalar. Also calculating curvatures, one finds
that the results can be presented in an (double) expansion in 
$r^2/\beta L^2$ (as well as $L^2/r^2$) in which the leading terms
correspond to those of the AdS geometry. 

However, with large but finite $\beta$, as we continue to increase
$r$ eventually we reach a regime where $r>\beta$ and the perturbative
expansion above in $r^2/\beta L^2$ breaks down. Essentially we have
again entered a nonlinear regime, now at large $r$,
 where the analysis of the linearized 
equations of motion is insufficient. In the
present case, however, we know precisely what the physics of this
nonlinear regime is since we have the full supergravity solutions
\reef{fullm}. In particular, the $\del\chi_+$ mode does not introduce
any divergent curvatures, but it does modify the asymptotic
structure of the solution from AdS to that of Minkowski space.
So for example in $R_{MN}R^{MN}$, the divergent series constructed
with the perturbative expansion above is resummed into the full
result which in fact vanishes as $r\rightarrow\infty$.

Thus we see that the irrelevant operator $\cO_8$ in the CFT controls
how the near-horizon geometry expands into asymptotically
flat space.
This discussion then extends the analysis of ref.~\cite{aki}
where they considered modifying the super-Yang-Mills Lagrangian
by the addition of $\cO_8$, and gave evidence that this
perturbation allowed one to match scattering cross-sections beyond the
near-horizon limit. Of course, this discussion in which we do not
fully implement the decoupling limit and how it may be related to
introducing a irrelevant operator in the CFT applies in general, and
not just to the particular supergravity solutions studied in this
paper.

\smallskip
\noindent{\bf (v) Consistency and singularities:}

Let us return to discussing the completely decoupled near-horizon solution,
and in particular we will consider the Lorentz invariant solution 
\reef{ds} which was the focus of our investigation. Despite the fact
that above we argued that this solution should interpreted in terms
of introducing expectation values for certain operators in the dual
gauge theory, we still ask whether the dual theory deserves to
be called a new ``phase'' of the super-Yang-Mills theory. On the one
hand, we have a Lorentz invariant state with {\it zero energy},
as indicated below eq.~\reef{noenergy}. However, we have nontrivial
expectation values, \eg $\langle\cO_4\rangle\propto\dee\w^4$, which
distinguishes this state from the conventional vacuum of the
super-Yang-Mills theory. There is no immediate contradiction between
the nonvanishing expectation value of $\langle\cO_4\rangle$
and $\langle T_{ab}\rangle=0$, even though the expectation values involve
linear combinations of the same operators. Because
these are normal-ordered composite operators whose expectation values
may be  positive or negative, delicate cancellations can occur
in $\langle T_{ab}\rangle$ and still leave $\langle\cO_4\rangle$
nonvanishing --- note that it is important that
$\eta^{ab}T_{ab}$ for the CFT in order to avoid a contradiction.
 It seems clear though that this Lorentz
invariant zero-energy state
can not be a part of the Hilbert space of states built on the conventional
{\it perturbative} vacuum.\footnote{We advocate this point of view
despite the recent discussion of
``precursors'' \cite{prec}, since the present system does not
involve  an localized excitations.}
Hence it would seem justified
in referring to the states dual to these supergravity
solutions as new phases of the theory. One may wonder if this new
phase represents an alternative quantization of the ${\cal N}=4$
super-Yang-Mills theory, and if it produces a fully consistent,
\eg stable, theory.

From the supergravity point of view, one may regard these solutions
as problematic since they contain naked time-like singularities.
Certainly
in many cases, string theory is able to resolve 
 singular classical geometries \cite{zing}.
One should keep in mind, however, that one does not expect
string theory to resolve all possible singularities \cite{winner}.
For example, one does not expect the negative energy Schwarzschild
solution (or its AdS cousin, in the present context) to play
a role in a fully consistent superstring theory. It would seem
that solutions studied here fall into the same category of
problematic solutions with time-like singularities 
and ``unusual'' energies. It could be that
there are various superselection sectors and that the singular 
solutions are part of a separate sector
of the theory, as occurs in the standard Kaluza-Klein framework
\cite{kalku}.

Also we note that given smooth nonsingular initial 
data in the supergravity
solution, an asymptotically AdS solution would be protected from
evolving into a solution resembling those considered here, by
the cosmic censorship conjecture \cite{censor}. For example, if one
set up a shell of infalling matter which at the same time carried
scalar hair, \eg dilaton or volume scalar excitations, one would
expect the hair to be radiated away and that the collapse would
produce a conventional black hole. This may be related to the fact
that we could not find evidence of a mass gap in the dilaton and
metric excitations of the dilaton and metric fields in section 4.
That is if the singularity is resolved by adding matter
near $r=0$, it could be that the solution is wildly unstable
to radiating away the additional scalar hair.
In any event, cosmic censorship would seem to argue in favor of the
present solutions representing some exotic new super selection
sector of the SYM theory. 

Understanding when classically singular background solutions represent
physically admissible configurations within string theory is an important and 
longstanding question. The AdS/CFT correspondence seems to present a new
arena in which to to pursue this issue. Of course, the present investigation 
does not offer an resolution. One may hope though that by examining our 
backgrounds in combination with other singular asymptotically AdS solutions,
some generic features may emerge to produce some new insights into this 
question. Certainly one feature that the present work reveals is that only
examining a limited set of the properties of the singular solutions may
mislead one into thinking that the singularity is not relevant for low
energy physics. Here we saw that while the mass spectrum of the RR scalar
could be reliably calculated, that the dilaton (and the metric) depends on a
resolution of the singularity. It would be interesting to examine excitations
of a more extensive list of refs.~\cite{sfet,gub,extra,added,tsyt}
to see whether or not they are all shielded from the singularity by a 
potential barrier. We expect generically one will not find such
shielding from the singularity \cite{donm,afraid} for all of the
fields (including the massive supergravity and string modes).

\vspace{1cm}
{\bf Acknowledgements}

This research was supported by NSERC of Canada and Fonds
FCAR du Qu\'ebec. 
We would like to acknowledge useful conversations with 
Steve Gubser, Simeon Hellerman, Don Marolf, Guy Moore, Amanda Peet,
Vipul Periwal, Joe Polchinski and Simon Ross.
RCM would also like to thank the Institute for Theoretical Physics
at UCSB for its hospitality
while this paper was being finalized. While at the ITP, RCM was supported
by NSF Grant PHY94-07194.


\begin{thebibliography}{99}


\bibitem 
{juan}{J.M.~Maldacena,
Adv. Theor. Math. Phys. {\bf 2}, 231 (1997) hep-th/9711200.}

\bibitem 
{gkp}{S.S.~Gubser, I.R.~Klebanov and A.M.~Polyakov,
Phys. Lett. {\bf B428}, 105 (1998)
hep-th/9802109.} 

\bibitem 
{ed}{E.~Witten,
Adv. Theor. Math. Phys. {\bf 2}, 253 (1998) hep-th/9802150.}

\bibitem 
{revue}{O.~Aharony, S.S.~Gubser, J.~Maldacena, H.~Ooguri and Y.~Oz,
{\it ``Large N field theories, string theory and gravity,''}
hep-th/9905111.

\bibitem 
{edd}{E.~Witten,
Adv. Theor. Math. Phys. {\bf 2}, 505 (1998) hep-th/9803131.}

\bibitem 
{poly}{A.M.~Polyakov, 
Int. J. Mod. Phys. {\bf A14}, 645 (1999) hep-th/9809057.}

\bibitem 
{ooguri}{C.~Csaki, H.~Ooguri, Y.~Oz and J.~Terning,
JHEP {\bf 01}, 017 (1999) hep-th/9806021;
H.~Ooguri, H.~Robins and J.~Tannenhauser,
Phys. Lett. {\bf B437}, 77 (1998) hep-th/9806171;
R.~de Mello Koch, A.~Jevicki, M.~Mihailescu and J.P.~Nunes,
Phys. Rev. {\bf D58}, 105009 (1998) hep-th/9806125;
M.~Zyskin,
Phys. Lett. {\bf B439}, 373 (1998) hep-th/9806128.}

\bibitem 
{mini}{J.A.~Minahan,
JHEP {\bf 01}, 020 (1999) hep-th/9811156.}

\bibitem 
{sfet3}{J.G.~Russo and K.~Sfetsos,
Adv.\ Theor.\ Math.\ Phys.\ {\bf 3}, 131 (1999)
hep-th/9901056; 
 C.~Csaki, J.~Russo, K.~Sfetsos and J.~Terning,
Phys.\ Rev.\ {\bf D60}, 044001 (1999)
hep-th/9902067.}

\bibitem 
{more}{J.G.~Russo,
Nucl. Phys. {\bf B543}, 183 (1999) hep-th/9808117;
C.~Csaki, Y.~Oz, J.~Russo and J.~Terning,
Phys. Rev. {\bf D59}, 065008 (1999) hep-th/9810186;
A.~Hashimoto and Y.~Oz,
Nucl.\ Phys.\ {\bf B548}, 167 (1999)
hep-th/9809106.}

\bibitem 
{li}{M.~Li, 
JHEP {\bf 9807} (1998) 014, hep-th/9804175;
JHEP {\bf 9807} (1998) 003, hep-th/9803252.}

\bibitem 
{gross}{D.~Gross and H.~Ooguri,
Phys.Rev. {\bf D58}, 106002 (1998) hep-th/9805129.}

\bibitem 
{zero}{I.R.~Klebanov and A.A.~Tseytlin,
Nucl.\ Phys.\ {\bf B546}, 155 (1999)
hep-th/9811035; 
Nucl.\ Phys.\ {\bf B547}, 143 (1999)
hep-th/9812089;
JHEP {\bf 03}, 015 (1999) hep-th/9901101.}

\bibitem 
{zero2}{G.~Ferretti and D.~Martelli,
Adv.\ Theor.\ Math.\ Phys.\ {\bf 3}, 119 (1999)
hep-th/9811208;
A.~Armoni, E.~Fuchs and J.~Sonnenschein,
JHEP {\bf 06}, 027 (1999)
hep-th/9903090;
M.~Alishahiha, A.~Brandhuber and Y.~Oz,
JHEP {\bf 05}, 024 (1999)
hep-th/9903186;
G.~Ferretti, J.~Kalkhinen and D.~Martelli,
{\it ``Non-Critical Type 0 Strings and Field Theory Duals,''}
hep-th/9904013.}

\bibitem 
{joe2} J.A.~Minahan,
JHEP {\bf 04}, 007 (1999)
hep-th/9902074.


\bibitem 
{sfet}{A.~Kehagias and K.~Sfetsos,
Phys.\ Lett.\ {\bf B454}, 270 (1999)
hep-th/9902125.}


\bibitem 
{gub}{S. S. Gubser {\it ``Dilaton Driven Confinement,''} hep-th/9902155.  }

\bibitem 
{extra}{A.~Kehagias and K.~Sfetsos,
Phys.\ Lett.\ {\bf B456}, 22 (1999)
hep-th/9903109; S.~Nojiri and S.D.~Odintsov,
Phys.\ Lett.\ {\bf B458}, 226 (1999) hep-th/9904036;
{\it ``Curvature dependence of running gauge coupling and confinement in
AdS/CFT correspondence,''} hep-th/9905200;
{\it ``Running gauge coupling and quark antiquark potential in non-SUSY gauge
theory at finite temperature from IIB SG/CFT correspondence,''}
hep-th/9906216.} 

\bibitem 
{added}{R.~de Mello Koch, A.~Paulin-Campbell and J.P.~Rodrigues,
{\it ``Non-Holomorphic Corrections from Threebranes in F-Theory''},
hep-th/9903029.
S.~Nojiri and S.D.~Odintsov, Phys. Lett. {\bf B449} (1999) 39, hep-th/9812017.}  }

\bibitem 
{tsyt}{H.~Liu and A.A.~Tseytlin,
Nucl.\ Phys.\ {\bf B553}, 231 (1999) hep-th/9903091.}

\bibitem 
{john}{J.H. Schwarz, Nucl. Phys. {\bf B226}, 269 (1983).} 

\bibitem 
{jaun2}{J.~Maldacena,
Phys.\ Rev.\ Lett.\ {\bf 80}, 4859 (1998)
hep-th/9803002.}

\bibitem 
{rey2}{S.J. Rey and J. Yee,
{\it ``Macroscopic Strings as Heavy Quarks of Large N Gauge Theory
and Anti-deSitter Supergravity,''} hep-th/9803001;
S.~Rey, S.~Theisen and J.~Yee,
Nucl.\ Phys.\ {\bf B527}, 171 (1998)
hep-th/9803135.}

\bibitem 
{itz}{A. Brandhuber, N. Itzhaki, J. Sonnenschein and S. Yankielowicz,
JHEP {\bf 06}, 001 (1998) hep-th/9803263.}

\bibitem 
{mono}{J.A.~Minahan,
{\it ``Quark - monopole potentials in large N superYang-Mills,"}
Adv. Theor. Math. Phys. {\bf 2}, 559 (1998)
hep-th/9803111.}

\bibitem 
{suss}{L. Susskind and E. Witten, {\it ``The Holographic Bound in
 Anti-de-Sitter Space,''} hep-th/9805114.  }

\bibitem 
{peet}{A.W.~Peet and J.~Polchinski,
Phys.\ Rev.\ {\bf D59}, 065011 (1999)
hep-th/9809022.}
 
\bibitem 
{donm}{G.T.~Horowitz and D.~Marolf,
Phys. Rev. {\bf D52}, 5670 (1995) gr-qc/9504028;
C.F.~Holzhey and F.~Wilczek,
Nucl. Phys. {\bf B380}, 447 (1992) hep-th/9202014.}

\bibitem 
{univ}{L.~Girardello, M.~Petrini, M.~Porrati and A.~Zaffaroni,
JHEP {\bf 05}, 026 (1999)
hep-th/9903026.}

\bibitem 
{sodd} see, for example: M.~Cvetic and C.M.~Hull,
Nucl. Phys. {\bf B480}, 296 (1996) hep-th/9606193.

\bibitem 
{cmp} C.G.~Callan, R.C.~Myers and M.J.~Perry, 
Nucl. Phys. {\bf B311}, 673 (1989).

\bibitem 
{russo} J.G.~Russo, Phys. Lett. {\bf B435}, 284 (1998)
hep-th/9804209;
Y.~Kiem and D.~Park, Phys. Rev. {\bf D59}, 044010 (1999) hep-th/9809174.

\bibitem 
{alba}{V. Balasubramanian, P. Kraus and A. Lawrence,
Phys. Rev. {\bf D59} (1999) 046003, hep-th/9805171;
V.~Balasubramanian, P.~Kraus, A.~Lawrence and S.P.~Trivedi,
Phys. Rev. {\bf D59}, 104021 (1999) hep-th/9808017.}

\bibitem 
{scan}{U.H.~Danielsson, E.~Keski-Vakkuri and M.~Kruczenski,
JHEP {\bf 01}, 002 (1999) hep-th/9812007.}

\bibitem 
{stressa} {R.C.~Myers,
{\it ``Stress tensors and Casimir energies in the AdS/CFT correspondence,"}
hep-th/9903203, to appear in Physical Review D.}

\bibitem 
{black} see, for example:
A.~Chamblin, R.~Emparan, C.V.~Johnson and R.C.~Myers,
``Holography, thermodynamics and fluctuations of charged AdS black holes,"
hep-th/9904197, to appear in Physical Review D.

\bibitem 
{ark} A.A.~Tseytlin,
Nucl. Phys. {\bf B501}, 41 (1997) hep-th/9701125.

\bibitem 
{stress}{V.~Balasubramanian and P.~Kraus,
{\it ``A Stress tensor for Anti-de Sitter gravity,"}
hep-th/9902121.}

\bibitem 
{freetown} P.~Breitenlohner and D.Z.~Freedman,
Phys. Lett. {\bf 115B}, 197 (1982);
Ann. Phys. {\bf 144}, 249 (1982);
L.~Mezincescu and P.K.~Townsend,
Phys. Lett. {\bf 148B}, 55 (1984);
Ann. Phys. {\bf 160}, 406 (1985).

\bibitem 
{renorm} D.Z.~Freedman, S.S.~Gubser, K.~Pilch and N.P.~Warner,
{\it ``Renormalization group flows from holography supersymmetry and a
c theorem,"} hep-th/9904017; {\it ``Continuous distributions of D3-branes
and gauged supergravity,''} hep-th/9906194; 
A.~Karch, D.~Lust and A.~Miemiec,
Phys.\ Lett.\ {\bf B454}, 265 (1999)
hep-th/9901041;
J.~Distler and F.~Zamora,
Adv.\ Theor.\ Math.\ Phys.\ {\bf 2}, 1405 (1999)
hep-th/9810206;
L.~Girardello, M.~Petrini, M.~Porrati and A.~Zaffaroni,
JHEP {\bf 12}, 022 (1998)
hep-th/9810126.

\bibitem 
{aki}{S.S.~Gubser, A.~Hashimoto, I.R.~Klebanov and M.~Krasnitz,
Nucl. Phys. {\bf B526}, 393 (1998) hep-th/9803023;
S.S.~Gubser and A.~Hashimoto,
Commun.\ Math.\ Phys.\ {\bf 203}, 325 (1999)
hep-th/9805140.}

\bibitem 
{prec} J.~Polchinski, L.~Susskind and N.~Toumbas,
``Negative energy, superluminosity and holography,"
hep-th/9903228.

\bibitem 
{zing}{see, for example: G.T.~Horowitz and A.R.~Steif,
Phys. Rev. Lett. {\bf 64}, 260 (1990);
C. Vafa, ``Strings and singularities,"
hep-th/9310069 in {\it Salamfestschrift: A collection of talks},
eds., A. Ali, J. Ellis and S. Randjbar-Daemi (World Scientific, 1993).}

\bibitem 
{winner}{G.T.~Horowitz and R.~Myers,
Gen. Rel. Grav. {\bf 27}, 915 (1995) gr-qc/9503062.}

\bibitem 
{kalku}E. Witten, Nucl. Phys. {\bf B195}, 481 (1982);
D. Brill and G. Horowitz, Phys. Lett. {\bf B262}, 437 (1991);
D. Brill and H. Pfister, Phys. Lett. {\bf B228}, 359 (1989).

\bibitem 
{censor}{for a recent review, see: R.M.~Wald,
``Gravitational collapse and cosmic censorship,"
gr-qc/9710068.}

\bibitem 
{afraid} A.~Ishibashi and A.~Hosoya,
{\it ``Who's afraid of naked singularities?
Probing timelike singularities  with finite energy waves,''}
gr-qc/9907009.


\end{thebibliography}
\end{document}